\documentclass[aps,prb,twocolumn,groupedaddress,floatfix,showpacs]{revtex4-2}

\usepackage{graphicx}
\usepackage{amsmath}
\usepackage{subcaption}
\usepackage[T1]{fontenc}
\usepackage{color}

\begin{document}


\title{Optical to microwave frequency conversion with Rydberg excitons}


\author{David Ziemkiewicz}
\email{david.ziemkiewicz@utp.edu.pl}

\author{Sylwia Zieli\'{n}ska-Raczy\'{n}ska}

 \affiliation{Institute of
Mathematics and Physics, Technical University of Bydgoszcz,
\\ Al. Prof. S. Kaliskiego 7, 85-789 Bydgoszcz, Poland}


\date{\today}

\begin{abstract} 
A novel, copper-based plasmonic system is presented to provide optical to microwave photon conversion. The process uses highly excited levels in Cu$_2$O Rydberg excitons and takes advantage of spoof plasmons, which allow for significant enhancement of the transition probability between specific excitonic energy levels. The theoretical results are verified with numerical simulations. The proposed system is very flexible, allowing for emission of microwaves wavelength  from 0.1 mm to 10 mm.   
\end{abstract}


\maketitle
\section{Introduction}
The process of  optical frequency conversion provides a fundamental and important approach to modify light in frequency domain. Possibilities of manipulating the frequency  without changing the information of other degrees of freedom of light will enable one to establish an interface between various optical systems operating in different frequency regions and have many classical and quantum applications \cite{Xan21,Lambert20}.
Photons play an important role in quantum information science because they are excellent carriers of quantum information. A variety of quantum qubits systems exploiting excitations at optical frequencies or  superconducting qubits  operating at microwave frequencies have been described and practically applied \cite{Zhong, Wendin}. However, optical photons can easily propagate over large distances while microwave photons can be easily controlled in a relatively macroscopic circuits \cite{Han2021}.
Hence, designing photonic transducers, an important factor, which has to be taken into account, is  an interplay between benefits and  disadvantages, inherent   their physical character. 
Different components for a synchronic photon generation and manipulation in practical systems cannot interact directly due to the wavelength and frequency mismatch, thus interfaces are needed to overcome such problems. In particular, a photon frequency interface capable of converting frequencies and bandwidths is indispensable.
Improved coupling between optical and microwave frequencies would enhance classical telecommunications and would be apply in distributed quantum networks and quantum communication. 

The first step towards many microwave applications is an efficient conversion of energy from optical frequencies to microwave ones. One of the techniques involves accelerating surface charge carriers in a semiconductor with femtosecond laser \cite{Chau2021,Ramanandan2014,Gupta2018}, but this method results in a very short emission time and correspondingly wide emission spectrum; for example, an emission spectrum ranging from 0.1 to 7.5 THz is reported in \cite{Ramanandan2014}. This limitation also applies to THz emission from structured metallic surfaces \cite{Oladyshkin2019} and photoconductor-based systems \cite{Auston84,Wang2022}. Our proposal can operate under continuous illumination and is tied to the spectral width of specific excitonic level transitions, which can vary from GHz to MHz \cite{DZ_OE}, making them significantly more narrowband. Another way of coupling is the use of Rydberg states of atoms placed in a cavity \cite{Gard17}, but this approach demands vacuum chamber and cryogenic cooling. In this paper, we propose the use of the newly discovered solid-state analogue of Rydberg atoms - Rydberg excitons in Cu$_2$O interacting with plasmons for the purpose of frequency conversion.

An exciton in a semiconductor is a quasiparticle consisting of conduction band electron and a valence band hole bound by the Coulomb interaction. As a whole, it is an electrically neutral particle with many properties similar to hydrogen atom. Rydberg excitons (REs) are the states characterized by a large principal quantum number and share many properties with Rydberg atoms. Excitonic states in Cu$_2$O with principal quantum number up to $n=$25 have been observed in 2014 \cite{Kazimierczuk}. Since then, this area of study develops rapidly; right now, many groups focus on exploring the properties of REs in low-dimensional systems \cite{Poddubny,DZ9,Hamid,Konzelmann} and fabrication techniques of Cu$_2$O nanostructures \cite{Steinhauer,Takahata2018}. Highly excited resonances in REs Cu$_2$O system are very close to each others, they are separated by several meV, which situates these transitions in microwave region and allows for controlled emission of microwaves \cite{Huber06}.
 This makes Cu$_2$O a very promising candidate for solid state microwave devices such as maser \cite{maser}. 
In 2018 we proposed a continuous-mode solid-state maser based on
Cu$_2$O, in which an ensemble of highly excited Rydberg exciton
states serves as a gain medium \cite{maser}. It was shown that the system
is highly tunable with an external electric field, allowing
for a wide range of emission frequencies within terahertz
range \cite{DZ_OE}.
 In contrast to these works, where a resonant cavity has been used as a device for enhancing the probability of specific microwave transitions, here we focus on plasmonic structures and  plasmon-exciton interactions due to their advantages in miniaturization and ease of fabrication\cite{DZ2022,Khurgin19}. Moreover, we do not rely on the nonlinear properties of the system, which is the basis of many conversion schemes \cite{Weber2018,Shen2021,Agreda2021}; these solutions tend to have relatively low efficiency, photon emission rate on the order of $10^4$ per second \cite{Shen2021} and need sufficient power to take advantage of the nonlinear properties.
 
Plasmons are combination of light and a collective oscillation of  the free electron plasma at metal-dielectric interface. Surface plasmon polaritons (SPPs) can be described as  dual  waves of charge motion in a metal and electromagnetic waves in the surrounding medium.
 In general, it has been demonstrated that one can efficiently couple plasmons with excitons, forming so-called plexcitons \cite{Fofang08,Karademir14}, which have been proposed as a basis of  variety of tunable devices \cite{Lee16,Cao18,Goncalves18}. Usually, only plasmons in IR-visible-UV range are considered, which are chosen to match the energy of specific excitonic states \cite{DZ2022}. However, for microwave inter-excitonic transitions, one has to take advantage of  the so-called spoof plasmons \cite{Pendry04}, which
 are surface electromagnetic waves in microwave and terahertz regimes, they propagate along planar interfaces with sign-changing permittivities. Since surface plasmons  cannot appear naturally in microwave and terahertz frequencies due to dispersion properties of metals, spoof surface plasmons require artificially-engineered materials \cite{Kong2015, Joy2017, Erementchouk16, Tang2018}. 
Finally, we note that Cu nanostructure can match the plasmon lifetime of the more traditional materials such as silver or gold \cite{Chan07,Mkhitaryan21}, especially when coupled to a controlled layer of oxide \cite{Rodriguez11,Parramon19}. Therefore, the use of Cu-Cu$_2$O nanostructure provides multiple advantages: direct coupling to Cu$_2$O excitons, low plasmonic energy loss and ease of fabrication \cite{Bohme19,Melo18,Huang20}. Moreover, the terahertz radiation can be either emitted as free space photons with the use of coupling structure or kept in the form of surface plasmons, which is particularly valuable option due to the lack of direct SPP sources in the THz regime \cite{Zhang2020}.

The paper is organized as follows. In the first section, the theoretical descriptions of Rydberg excitons features, Purcell effect and spoof plasmons are outlined. Then, the possibility of enhancing specific transition probabilities is analysed theoretically and numerically and the optimal system geometry is indicated. In the next section, the amplification factors are used to calculate the microwave emission power for specific excitonic transitions and to estimate conversion efficiency. The possibility of single photon level emission is also discussed. Finally, the conclusions are presented. The details of the numerical method used for field calculations are outlined in the Appendix A.

\section{Theory}
\subsection{Rydberg excitons and Rydberg blockade}
 Rydberg exciton is a highly excited electron-hole pair bound
by Coulomb attraction, characterized by a principal quantum number $n$ >>1 for dipole-allowed
P-type envelope wavefunctions. In principle, REs have similar scaling as Rydberg atoms 
 although the physical origin of these similarities in their case bases on a complex valence band
structure and different selection rules for excitons. The dimensions of REs scale as $n^2$ and can reach micrometers, and their life-times are proportional to $n^3$ and can reach hundreds of nanoseconds \cite{Kazimierczuk}.
 Another exceptional characteristic is their strong interaction with external fields due to the fact that their polarizability scales as $n^7$.
In REs systems in copper oxide electric dipole transitions in the microwave frequencies region (10-1000 GHz) can be accessed with $n+1 \rightarrow n$ transitions, for $n>8$. Interactions involving coupling of REs with microwaves  have been observed recently in Cu$_2$O by Gallagher \textit{et al} \cite{Gallagher}  which confirms that the system of copper oxide with REs is a very promising candidate for solid state microwave devices such as maser \cite{maser}.  
  
 Significant feature of Rydberg excitons is Rydberg blockade, which
 results of a long-range dipole-dipole and Van der Waals interactions between them, which can
 be large enough to perturb the energy level of nearby excitons, so they no longer have the same frequency, which prevents their excitation in the immediate vicinity of already existing exciton.
 The space where another exciton cannot be created is described by a so-called blockade volume, which scales extremly fast as $n^7$ \cite{Kazimierczuk}
\begin{equation}\label{e_blok}
V_b \approx 3 \cdot 10^{-7} \mu m^3 \cdot n^7.
\end{equation}
As a consequence, highly excited states very quickly reach the saturation level when the medium is completely filled with excitons. Thus, the light propagating through the medium is not
absorbed to create new ones. In conclusion, the maximum number of excitons that can exist in any given volume is strictly limited. In the system presented below, this volume is the immediate vicinity of the proposed plasmonic structure. As it will be shown later, aside from thermal considerations, Rydberg blockade is the main limiting factor on the microwave emission power.

\subsection{Material model and system setup}
In our previous papers describing the  maser on Rydberg excitons, we proposed the use of a metallic cavity for enhancing the probability of microwave transitions \cite{maser}. However, such an approach has some limitations. The Cu$_2$O crystal that fills the cavity becomes inconveniently large in the case of longer wavelengths approaching 1 mm; for such a size, it becomes opaque for visible light, so  an illuminating optical beam penetrates only the surface layer of the medium, up to the depth of few tens of $\mu$m. Thus only a small fraction of the crystal volume is occupied by excitons, which is an important limitation considering the large volume taken by excitons and the resulting Rydberg blockade.

An alternative approach, which allows to overcome this limitation, is some sort of resonant, plasmonic structure that is surrounded by Cu$_2$O. The first plasmonic structure capable of sustaining microwave plasmons has been proposed by Pendry \cite{Pendry98} and had a form of thin, metallic wires. We recall that the electron plasma frequency is given by 
\begin{equation}\label{e_wp}
\omega_p^2 = \frac{Ne^2}{\epsilon_0m_e},
\end{equation}
where $N$ - is the charge carrier (electron) concentration, $m_e$ is the electron effective mass. For copper, the plasma frequency $\hbar\omega_p \approx 7.4$ eV, which corresponds to the ultraviolet spectral range. However, as shown by Pendry \cite{Pendry98}, the effective mass of electron can be greatly enhanced in a sufficiently thin wire, resulting in decreased plasma frequency. This effect can be attributed to the fact that an electron moving in a very thin wire generates significant magnetic field in the limit of the wire radius $r \rightarrow 0$, which increases the inductance of the system. That additional inductance disturbs the acceleration of the electron and thus increases its apparent inertia, i.e. the effective mass, allowing for efficient coupling to microwave frequency radiation unlike the optical frequencies usually employed in plasmonics. This particular mechanism has been used to facilitate negative permittivity in the first experimentally demonstrated negative index metamaterials \cite{Shelby01}. The type of microwave surface wave used in this experiment is called a spoof plasmon \cite{Pendry04}. Since their discovery, the theory of spoof plasmons has been developing rapidly \cite{Rusina10}, many structures have been proposed \cite{Shen2014, Kong2015, Joy2017, Tang2018}, including recent applications in sensors \cite{Wang22}. 

The amplification of the emission rate of selected transitions has been first demonstrated by Purcell in a system consisting of Rydberg atom in resonant cavity \cite{Purcell,Todorov07}. Here, we take advantage of this phenomenon in a very similar context for Rydberg excitons. The amplification of emission rate is described by a Purcell factor \cite{Colas16}
\begin{equation}\label{eq:purcell}
F = \frac{3Q}{4\pi^2 V}\left(\frac{\lambda}{n}\right)^3,
\end{equation}
where $Q$ is the quality factor, V is effective mode volume (for a cavity, it is the cavity volume), $\lambda/n$ is the wavelength in the medium with refraction index $n$. In the case of plasmonic structure, the mode volume can be defined as \cite{Koendernick10}
\begin{equation}\label{V_eff}
V = \int \frac{\rho}{max(\rho)}d^3r,
\end{equation}
where $\rho$ is the electromagnetic energy density given by \cite{Landau,Nunes11}
\begin{equation}\label{eq:density}
\rho = \left(\epsilon + \omega\frac{\partial \epsilon}{\partial\omega}\right)|E|^2.
\end{equation}
The second term in parenthesis ensures that the energy density is positive in media with negative $\epsilon$. The effective volume $V$ is equal to the geometrical volume taken by the field in the case of constant field and smaller than geometrical volume when the field is strongly localized (like in the case of plasmons \cite{Palma12}). In contrast to the case of a cavity, the integration volume in Eq. (\ref{V_eff}) is not strictly defined, but only the space, where $\rho$ (and thus the electric field) is significant, can be considered. The quality factor of the plasmonic structure is defined as
\begin{equation}
Q = 2\pi\frac{E}{\Delta E},
\end{equation}
where $E$ is the total electromagnetic field energy contained in the system and $\Delta E$ is energy lost in a single period of the field oscillation.

It should be mentioned that the above definition of an effective volume $V$ is only a rough approximation \cite{Lambert20,Koendernick10} and several other approaches can be used to calculate the Purcell factor of a plasmonic system \cite{Simovski12,Poddubny13}. A detailed calculation of emission enhancement would demand more  sophisticated definition of an effective volume and inclusion of some secondary factors such as surface roughness and energy shift of excitonic levels in the vicinity of metal-dielectric interface. 

\begin{figure}
\includegraphics[width=.75\linewidth]{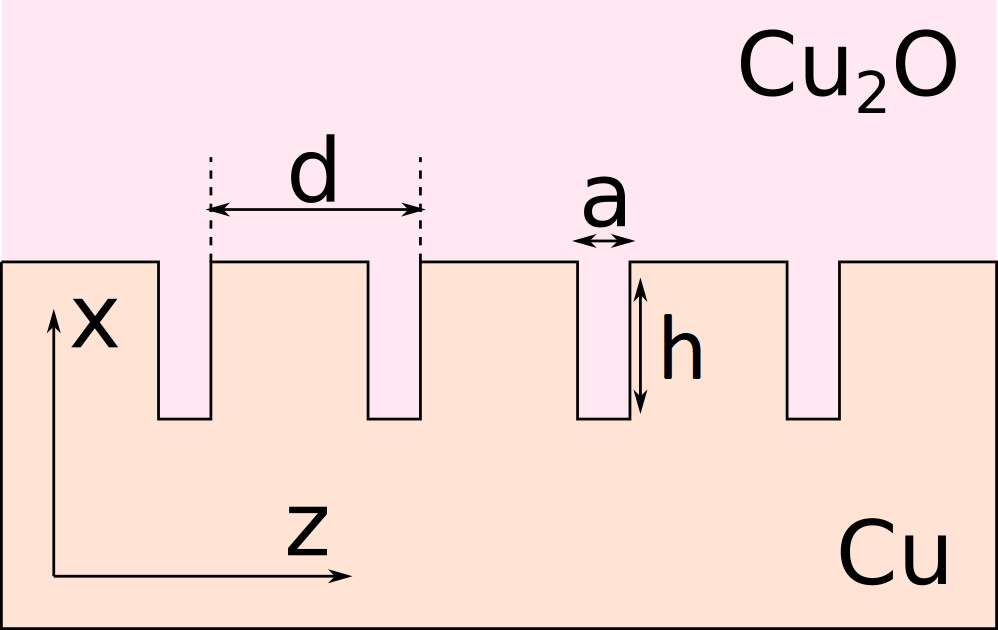}
\caption{Schematic representation of the system}\label{Fig_1}
\end{figure}

Let's consider a corrugated metal surface with multiple grooves with the depth $h$, width $a$ and period $d$ (Fig. \ref{Fig_1}). In this paper, we consider periods $d$ matched to the emission wavelength $\lambda \in (0.1-10)$ mm and $a,d \sim \lambda/10$.
The metal (Cu) is characterized by permittivity $\epsilon_m$, and the medium inside the grooves ($\epsilon_g$) and above them ($\epsilon_d$) are both Cu$_2$O. In the considered frequency range $\omega \sim 0.3$ THz one can assume $\epsilon_g=\epsilon_d \approx 7.5$, although it can vary depending on sample quality and temperature \cite{Weichman}. For metal, in this frequency range one has $\epsilon_m = -6218 + 90i$ \cite{Tukmakova20}. Note, that the extremely large negative permittivity of copper means that it is a good approximation of a perfect conductor, reflecting almost 100$\%$ of incoming radiation. The effective plasma frequency of the system depicted in Fig. \ref{Fig_1} is described by the formula $\omega_{p,eff} = \frac{\pi c}{2 h \sqrt{\epsilon_g}}$ \cite{Joy2017} and thus, in contrast to Eq. (\ref{e_wp}), it becomes dependent on the system geometry (distance $h$). This allows one to reduce the plasma frequency to the THz range.

The dispersion relation, which is derived in  \cite{Rusina10}, has the form
\begin{equation}
k=\left(\epsilon_dk_0^2 + \left[\frac{a\epsilon_d}{d\epsilon_g}\right]^2k_g^2\tan^2(k_gh)\right)^{0.5},
\end{equation}
where $k_0=2\pi f/c$ is the vacuum wave vector and
\begin{equation}
k_g = k_0 \sqrt{\epsilon_g} \left(1+\frac{l_s(i+1)}{a}\right),
\end{equation}
with $l_s=k_0 Re(\sqrt{-\epsilon_m})$ is the skin depth of the order of 30-120 nm in THz range \cite{Rusina10}. As mentioned above, in the microwave frequency range the copper can be treated as a perfect conductor. In such a case, it is possible to analytically calculate the field above the corrugated surface, provided that the grooves are considerably narrower than the wavelength \cite{Kong2015}. 

First, we define the effective SPPs refraction index as
\begin{equation}
n_{eff}=\frac{k_z}{k_0},
\end{equation}
where
\begin{equation}
\sqrt{n_{eff}^2-1}=\frac{a}{d}\tan(k_0h).
\end{equation}
Notably, the effective index, and thus optical properties, are  dependent only on geometric parameters of the structure ($a,d,h$).
Thus, we have
\begin{equation} \label{eq:kz}
k_z = k_0 \sqrt{1 + \frac{a}{d}\tan(k_0h)},
\end{equation}
and
\begin{equation}
k_x=\sqrt{k_z^2-k_0^2}.
\end{equation}
With the above values of wave vector components, one can use the Maxwell's equations to derive the expressions describing the EM field of the spoof plasmon, which have the following forms \cite{Kong2015}
\begin{eqnarray}\label{eq:fields}
H_y &=& H_0 e^{-k_x x}e^{i(k_z z-\omega t)},\nonumber\\
E_x &=& \frac{n_{eff}}{\epsilon_0 c}H_y, \nonumber\\
E_z &=& \frac{\sqrt{n_{eff}^2-1}}{i\epsilon_0 c}H_y,
\end{eqnarray}
where $H_0$ is a constant  dependent on source amplitude. Having the formula for the fields, we can use Eq. (\ref{eq:density}) to calculate the energy density and then effective volume (Eq. (\ref{V_eff})) and Purcell factor given by Eq. (\ref{eq:purcell}).

It should be stressed that the direct effect of excitons on the optical properties of Cu$_2$O is negligible; the change of permittivity due to excitonic resonance is on the order of $\Delta \epsilon \sim 10^{-3}$, which is very small when compared with $\epsilon_b=7.5$ \cite{DZ2022}. Therefore, the excitonic contribution can be ignored in the calculation of the electric field distribution. The role of excitons is to provide suitable energy levels to realize microwave transition that is enhanced by Purcell factor of the cavity.

\section{Field amplification with spoof plasmons}
\begin{figure}[ht!]
\centering
\includegraphics[width=.8\linewidth]{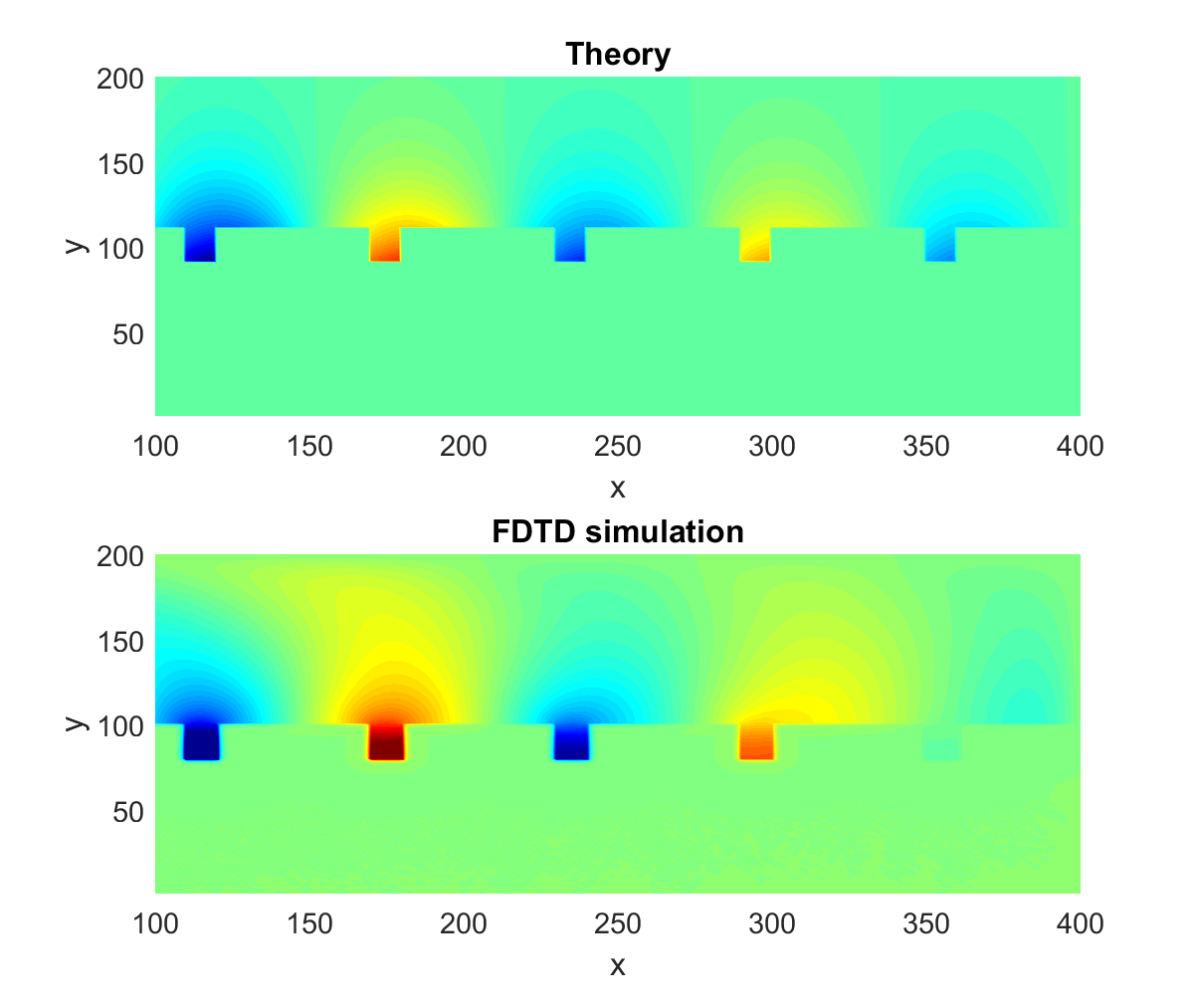}
\caption{Normalized field amplitude (color) obtained from Eq. (\ref{eq:fields}) and FDTD simulation, calculated for a structure characterized by $h=120$ $\mu$m, $a=80$ $\mu$m, $d=500$ $\mu$m.}\label{Fig_2}
\end{figure}

 Fig. \ref{Fig_2} shows a comparison between the field distribution obtained analytically from Eq. (\ref{eq:fields}) and in Finite-Difference Time-Domain (FDTD) simulation, the details of which are described in Appendix A. 
The system is characterized by groove depth $h=120$ $\mu$m, width $a=80$ $\mu$m and period $d=500$ $\mu$m. The operating wavelength  is $\lambda \approx 1$ mm. It should be noted that Eqs. (\ref{eq:fields}) describe only the field above the grooved surface \cite{Kong2015}, assuming that the corrugated metal layer below is a continuous, effective medium. For this reason, while the field on Fig. \ref{Fig_2} a) can be extended to the region inside the grooves, its values are not very accurate here. In fact, one can see that the field amplitude is somewhat greater on Fig. \ref{Fig_2} b). However, the good match between the field distributions at the top indicates that the numerical results are accurate. The obtained results are also consistent with similar simulations performed by others authors \cite{Kong2015,Erementchouk16,Joy2017}. One can also note that in the theoretical distribution of the field, the metal is a perfect conductor, with no field inside. On the other hand, in FDTD simulation the field enters the metal to a finite skin depth. Due to the limited spatial resolution $\Delta x = 10\mu$m, one cannot accurately measure that depth as it is smaller than a single unit cell.

\begin{figure}[ht!]
\centering
\includegraphics[width=.8\linewidth]{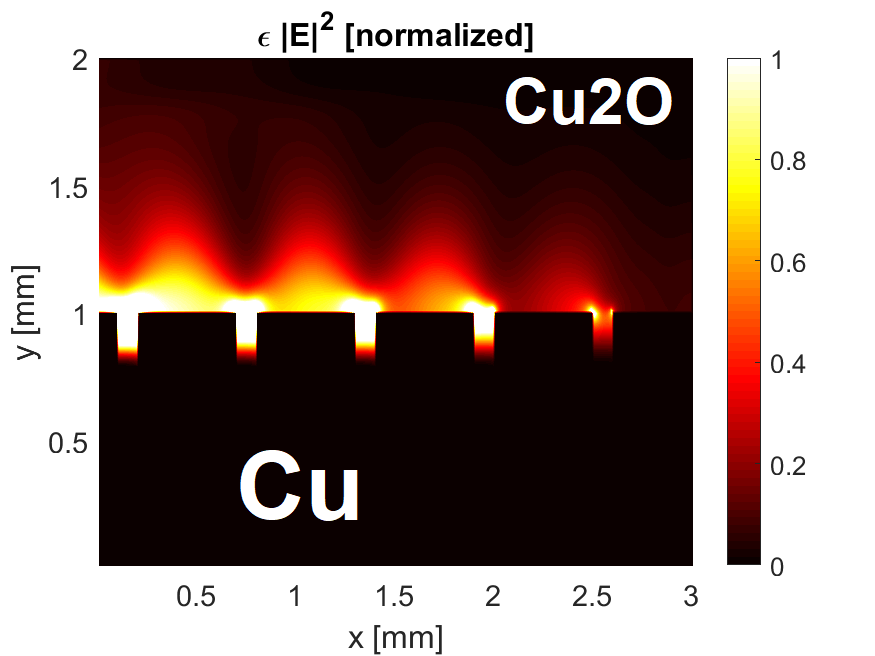}
\caption{Normalized energy density of the corrugated surface structure illuminated by $\lambda=1$ mm light; the same geometry as in Fig. \ref{Fig_2}.}\label{Fig_3}
\end{figure}

The important property of the system is the fact that the field is focused into deep subwavelength regions inside the grooves. In particular,  Fig. \ref{Fig_3} shows the time-averaged normalized field energy density in the system obtained in FDTD simulation. 
The concentration of energy in subwavelength spots greatly reduces the effective mode volume given by Eq. (\ref{V_eff}); for the discussed system, one obtains $V \approx 0.018 V_g$, where $V_g=\lambda^3$ is the geometric volume of the cavity. This result is comparable to the $V = 0.015 V_g$ reported in \cite{Maier06} for a metal-dielectric-metal system. For further comparison, an effective volume of $0.01$ $V_g$ is obtained in plasmonic particles \cite{Koendernick10} and up to $10^{-6}$ in a bow-tie antenna \cite{Yang16}. Overall, the grooved surface (also known as a nanoslit cavity) offers very small mode volume and relatively high $Q$ factor as compared to other plasmonic structures \cite{Chen18}. 

The $Q$ factor is strongly dependent on the geometry details; usually, it is of the order of $10^2$ \cite{Maier06,Wang22}, but with introduction of "defects" in the form of single shorter/longer grooves it can reach $10^5$ \cite{Joy2017}. Moreover, at microwave frequencies it is relatively straightforward to introduce gain to the system, increasing the $Q$ factor to $10^5$ \cite{Cai2018}. For the simulated structure shown in Fig. \ref{Fig_2}, the $Q$ factor has been estimated at $Q=124$. While the $Q$ factor is considerably lower than in regular cavities, the low mode volume still allows for large Purcell factor \cite{Colas16}.

With this in mind, we can calculate the Purcell factor from Eq. (\ref{eq:purcell}), obtaining $F \sim 530$. Again, the result is consistent with Purcell factors obtained in visible spectrum  of plasmonic structures; in \cite{Koendernick10}, $F \sim 10^2 - 10^5$ is reported in nanospheres depending on sphere diameter. The value of $F \approx 2000$ is demonstrated  for Ag nanocubes  \cite{Akselrod14}.  Maier \emph{et al}  reported  $F \sim 3400$  for the optical spectrum \cite{Maier06}, mentioning that for far infrared wavelengths  values exceeding $10^4$ are possible. For  metamaterial-dielectric interfaces similar conclusions are reported \cite{Ivanov20}. For the microwave frequency range an analysis of Purcell factor of a thin wire type plasmonic material yields $F \sim 10^2-10^3$ and up to $10^6$ for radio frequencies \cite{Poddubny12}.  

To investigate the potential of the corrugated metallic surface in order to improve the probability of microwave transitions, one first needs  to optimize its geometry. Again, we consider the structure from Fig. \ref{Fig_1} and perform FDTD simulations to obtain the EM field distribution. The simulations are two-dimensional, which implies that the system is large (longer than wavelength) in the $y$ axis. However, as noted in \cite{Tang2018}, the properties of spoof plasmons on such a structure depend very weakly on the thickness in $y$ axis, so that the results are applicable even to 10 nm thin films. This will be an important point later when Rydberg blockade is discussed. Moreover, we set the structure period fixed at $d=1$ mm, which is equal to the operating wavelength. In simulations, this value provided the best results regardless of other system dimensions; the match to the wavelength facilitates the formation of strong standing wave type plasmons with high field intensity. The Purcell factor $F$ and the structure $Q$ factor as a function of the groove depth $h$ and width $a$ are shown on Fig. 4.
\begin{figure}[ht!]
a)\includegraphics[width=.8\linewidth]{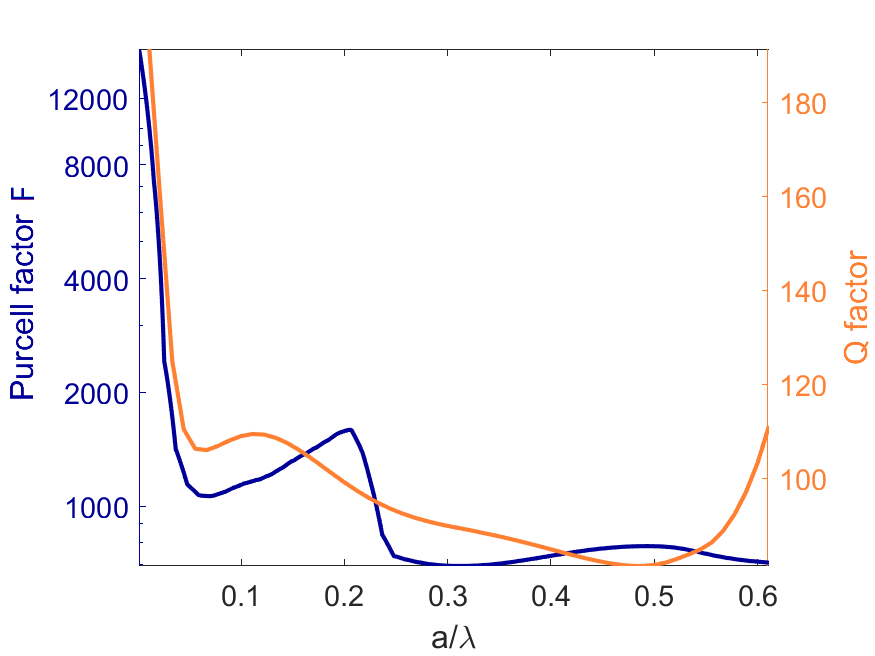}
b)\includegraphics[width=.8\linewidth]{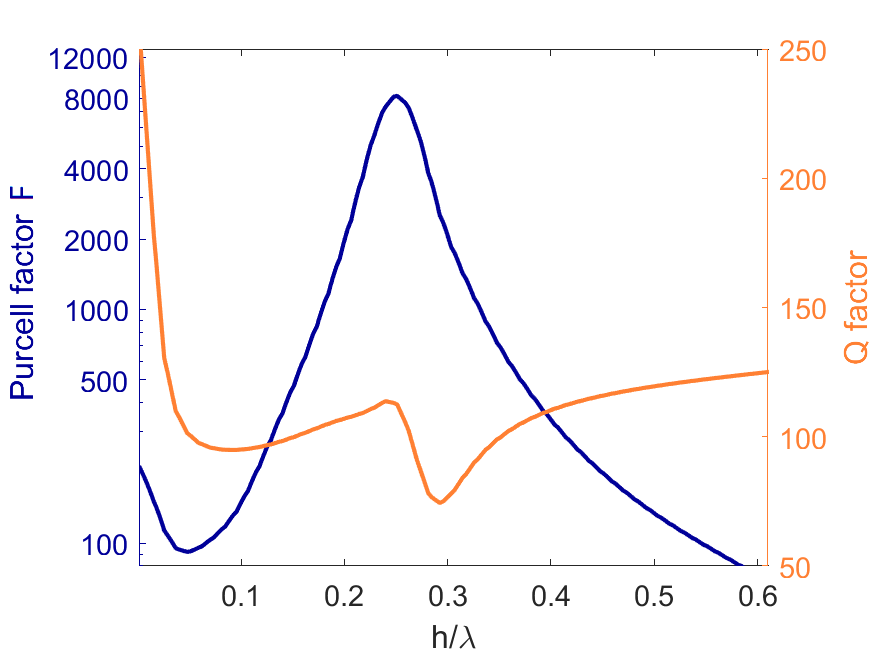}
\caption{Purcell factor $F$ and $Q$ factor of the structure as a function of a) width $a$ and b) depth $h$. Structure is illuminated by $\lambda=1$ mm light.}\label{Fig_4}
\end{figure} 
One can see on Fig. \ref{Fig_4} a) that the system has several distinct regimes of operation; in the limit of a very narrow groove, the field inside is highly focused, resulting in a small effective volume and considerable Purcell factor. Then, a second  local maximum occurs at $a \approx 0.2$ $\lambda$. As it will be shown later, depending on the groove height $h$, this maximum is located at $0.2$ - $0.3$ $\lambda$ and in general corresponds to the system where the groove acts as a quarter-wave cavity. Finally, a third, small maximum can be seen at $a \approx 0.5$ $\lambda$, which again facilitates the formation of standing waves. 

The quality factor  of the structure is approximately $Q \sim 100$ and shows relatively little variation with $a$, with exception of a very narrow grooves. However, from the simulation results it is inconclusive if the increase of $Q$ is the feature of narrow-gap SPP or the result of weaker coupling of the free space field to the plasmons; the stray waves that freely propagate through Cu$_2$O above the corrugated surface are very weakly attenuated and as such they spuriously increase the estimated $Q$ factor calculated from the propagation length. We stress that the value $Q$ serves only as a general estimation and is affected by many factors such as surface roughness that are omitted here.

The dependence of the $F$ on the groove height shown on  Fig. \ref{Fig_4} b) exhibits one, large local maximum at $h = 0.25$ $\lambda$. As noticed in \cite{Rusina10}, this is the expected result; the most efficient focusing of the field occurs when the grooves act as quarter-wave antennas. Interestingly, the $Q$ factor exhibits a local drop at this point. As noted in \cite{Erementchouk16}, well-defined SPPs are only possible when $h>d/2$. In our particular case, when $d=\lambda/2$, the limit is $h>\lambda/4$, which is exactly the point where the drop in Q factor occurs. This means that as $h$ increases, the system transitions from a mix of surface plasmons and free propagating waves to one where the energy located mostly in surface plasmons. As a result, the propagation distance is shortened and the $Q$ factor is reduced.

\begin{figure}[ht!]
\includegraphics[width=.8\linewidth]{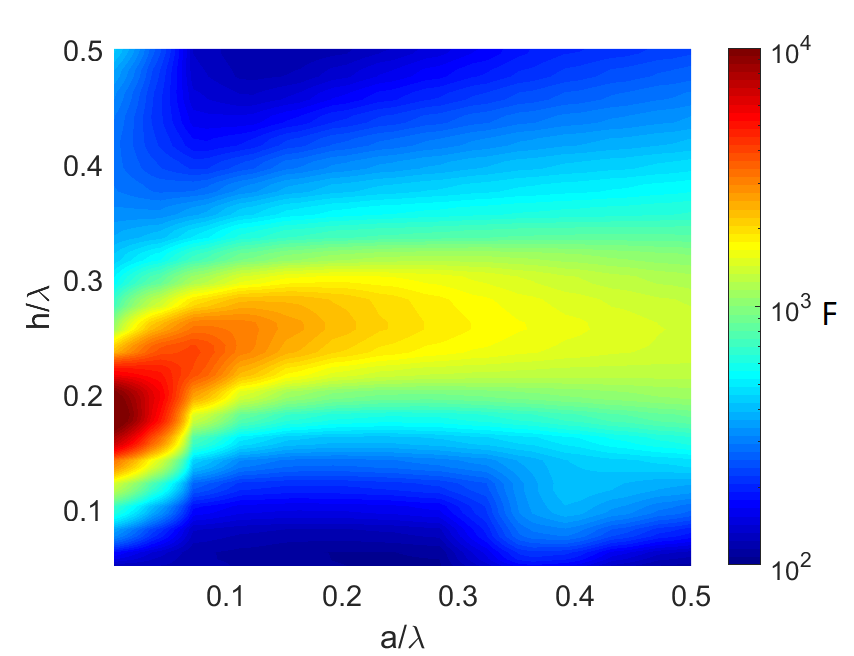}
\caption{Purcell factor $F$ of the structure as a function of $a$ and $h$, $\lambda=1$ mm.}\label{Fig_5}
\end{figure}

The optimization performed over full parameter space of $a$ and $h$ is shown on  Fig. \ref{Fig_5}. 
The main feature of the figure is a wide maximum that corresponds $h=0.25$ $\lambda$ for most values of $a$ and approaches $h=0.2$ $\lambda$ for $a \rightarrow 0$. As mentioned above, in general the Purcell reaches the maximum value for $h=0.25$ $\lambda$. However, for a very narrow grooves, the maximum shifts towards $h=0.2$ $\lambda$ due to the fact that the wave vector $k_z$ given by the dispersion relation  given by Eq. (\ref{eq:kz}) differs considerably from the free space wave vector $k_{z0}$ in the limit if $a \rightarrow 0$. In such a case, the wavelength is no longer equal to the nominal value of 1 mm and so the system is detuned from the chosen microwave transition.

The system geometry determines the upper limit of the emission power. For example, with $\lambda=1$ mm, $h=0.25\lambda$, $a=0.2\lambda$, the total geometrical volume of the cavity is $V_g=5 \cdot 10^7$ $\mu$m$^3$. Assuming that we use, for example, $n=10$ excitonic level, from Eq. (\ref{e_blok}) one has $V_b = 3 \mu$m$^3$ and thus the number of excitons in the cavity is $N \sim 3 \cdot 10^6$. This number, in conjunction with Purcell effect-boosted emission rate, allows one to estimate the number of emitted photons.

\section{Enhanced microwave emission}
\begin{figure}[ht!]
a)\includegraphics[width=.7\linewidth]{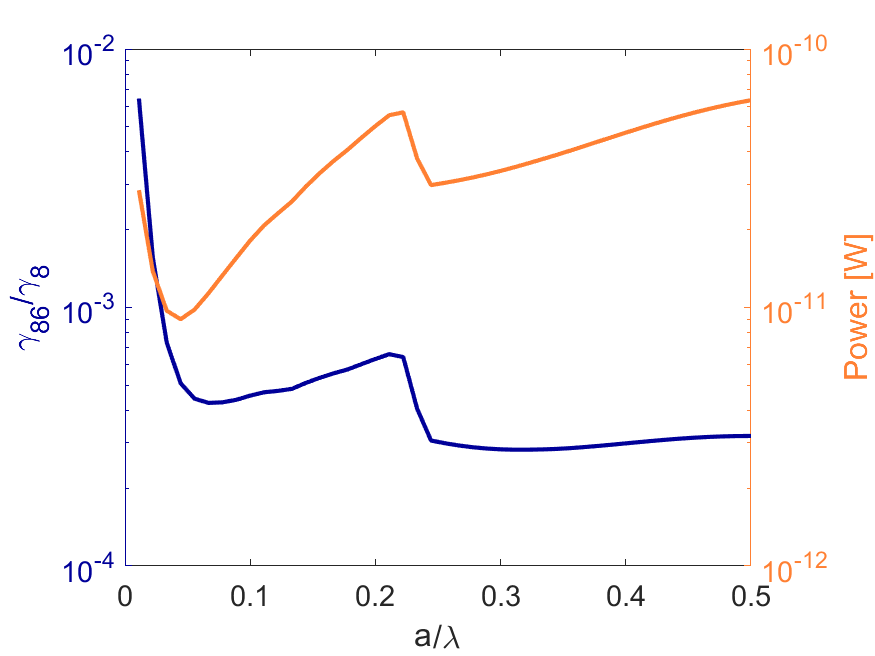}
b)\includegraphics[width=.7\linewidth]{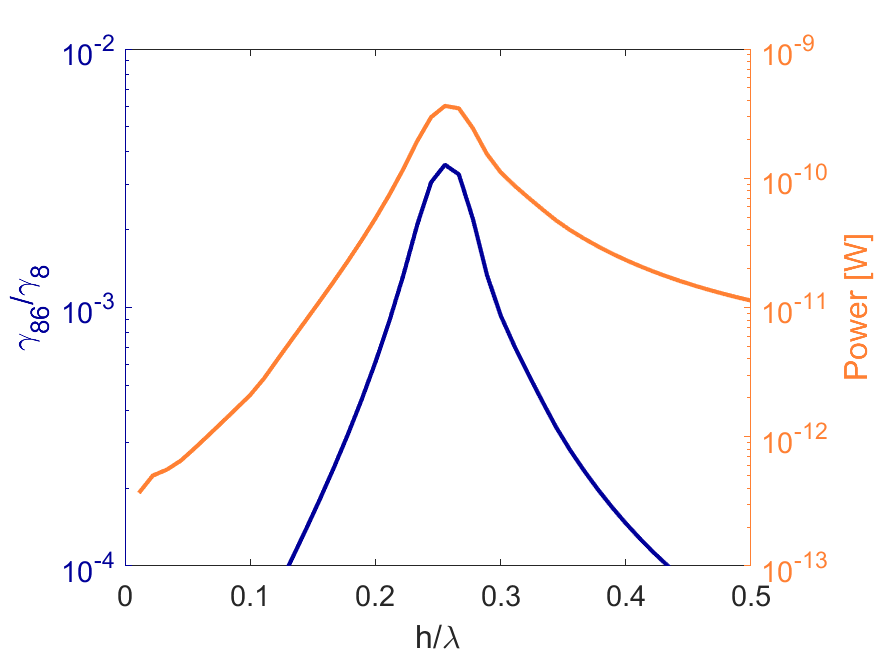}
\caption{The ratio of the transition rates $\gamma_{86}/\gamma_8$ and the microwave emission power as a function of a) $a$ and b) $h$; $\lambda=1$ mm.}\label{Fig_6}
\end{figure}

\begin{figure}[ht!]
a)\includegraphics[width=.8\linewidth]{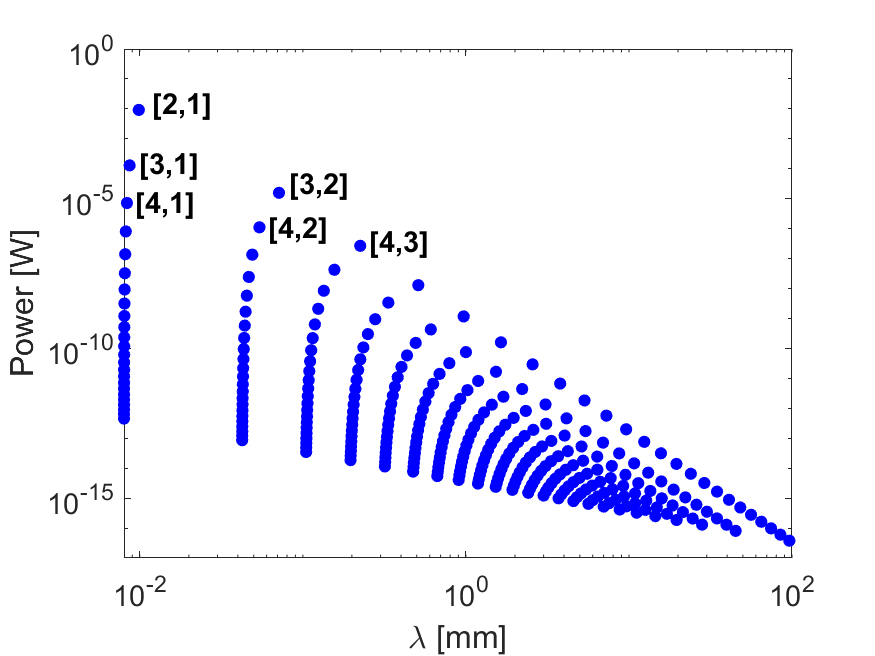}
\caption{The maximum microwave emission power as a function of wavelength; excitonic state combinations [$n_1,n_2$]; in each system $a = 0.2\lambda$, $h=0.25\lambda$, $d=0.5\lambda$. }\label{Fig_6b}
\end{figure}

With the results presented in the last section we can conclude that the Purcell factor $F \sim 10^5$ is possible in the discussed system. Now, we can investigate how such enhancement affects the exciton population dynamics in the Cu$_2$O. To perform such consideration let's assume that the system is illuminated by an light beam with frequency matched to the energy of $n=8$ exciton (2170.6 meV, 571 nm). The light intensity is sufficient to saturate the system, i.e., create the exciton density that approaches the limit imposed by Rydberg blockade; a single $n=8$ exciton is characterized by a blockade volume $V_B \approx 0.63$ $\mu$m$^3$ and so the upper limit of exciton density is $N_{max} \approx 1.6 \cdot 10^{12}$ cm$^{-3}$. The  wavelength for transition $n=8 \rightarrow n=6$ is $\lambda_{86} \approx 1.06$ mm is a close match to the idealized 1 mm wavelength discussed earlier. The point of interest is the transition rate $\gamma_{86}$ and how it compares to the most probable transition of the $n=8$ exciton back to the ground state, described by $\gamma_8 \approx 12.3$ GHz, $\hbar\gamma_8 \approx 25$ $\mu$eV \cite{Kazimierczuk, maser}. The rates are calculated on the basis of the overlap of hydrogen-like wavefunctions of $n=8$ and $n=6$ states, as described in detail in \cite{maser, DZ_OE} and result in $\gamma_{86} \approx 9.8$ kHz. The lower state relaxation rate is $\gamma_6 \approx 24.2$ GHz, $\hbar\gamma_8 \approx 36.7$ $\mu$eV. The results for a system with $h=0.25$ $\lambda$ are shown on  Fig. \ref{Fig_6} a).
As expected, in the limit of a narrow groove, the large Purcell factor results in the highest transition rate $\gamma_{86}$. In this regime, the fraction $\gamma_{86}/\gamma_8$ approaches $1\%$ which means that on average, for every 100 decaying $n=8$ excitons, one of them will return to the ground state through the intermediate state $n=6$, emitting microwave photon in the process. However, the emission power is proportional to the number of created excitons $N \sim \frac{V_g}{V_b}$ and thus it is relatively small in the case of a very narrow cavity with small volume $V_g$. Like in the earlier result in Fig. \ref{Fig_4}, the optimal operating point for power is the local maximum of Purcell factor at $a \sim 0.2$~$\lambda$ (Fig. \ref{Fig_6} a)). Similar optimization can be performed for the parameter $h$. In this case $h=\lambda/4$ is a global maximum for both power and transition rate (Fig. \ref{Fig_6} b)). This general characteristic is maintained in systems with different combinations of excitonic states. A full overview is presented on Fig. \ref{Fig_6b}.
The calculations have been performed for pairs of all states  [$n_1,n_2$],  $n_1 > n_2$ up to $n_1 = 25$. The emission power is given by
\begin{equation}
P = N\hbar\omega_{n_1,n_2}\gamma_{n_1,n_2}F.
\end{equation}
where $N$ is the average number of excitons inside the groove.
The aim is to illustrate some general tendencies and orders of magnitude and thus the quality factor is set to $Q=100$ and Purcell factor is assumed to be $F=10^4$. The plasmonic system is characterized by $a = 0.2$~$\lambda_{n_1,n_2}$ and $h=0.25\lambda_{n_1,n_2}$. A more thorough analysis of any particular state combination would warrant full geometry optimization for the given transition wavelength $\lambda_{n_1,n_2}$. The illuminating optical beam is assumed to be strong enough to saturate the system with excitons up to the limit imposed by Rydberg blockade.
One can notice that the emission wavelength on Fig. \ref{Fig_6b} is the most dependent on the choice of the lower excitonic level, with much smaller contribution from the upper level. This is a direct consequence of the energy spacing of excitonic levels $\delta E \sim n^{-1}$. The power decreases very quickly with $n$ due to several factors; the number of excitons in the system is inversely proportional to blockade volume of the excitons created by optical field, which scales as $n_1^7$. Moreover, the longer wavelength photons have smaller energy, which introduces another scaling factor $\sim \lambda_{n_1n_2}^{-1}$. The peak power is $10^{-2}$ W; however, it should be noted that the conversion efficiency is of the order of $10^{-2}$ and the ratio of optical photon to microwave photon energy is $\sim 10^3$ so overall, the maximum predicted power output would demand $P \sim 10^{3}$ W of an optical power, which could be only feasibly realized in pulsed operation. In general, the obtained power and efficiency figures are superior to approaches using nonlinear processes and comparable to the recently proposed plasmon-coupled surface state device \cite{Turan2021}. The highest obtained power corresponds to the photon emission rate in excess of $10^{16}$ per second. 

\begin{figure}[ht!]
a)\includegraphics[width=.8\linewidth]{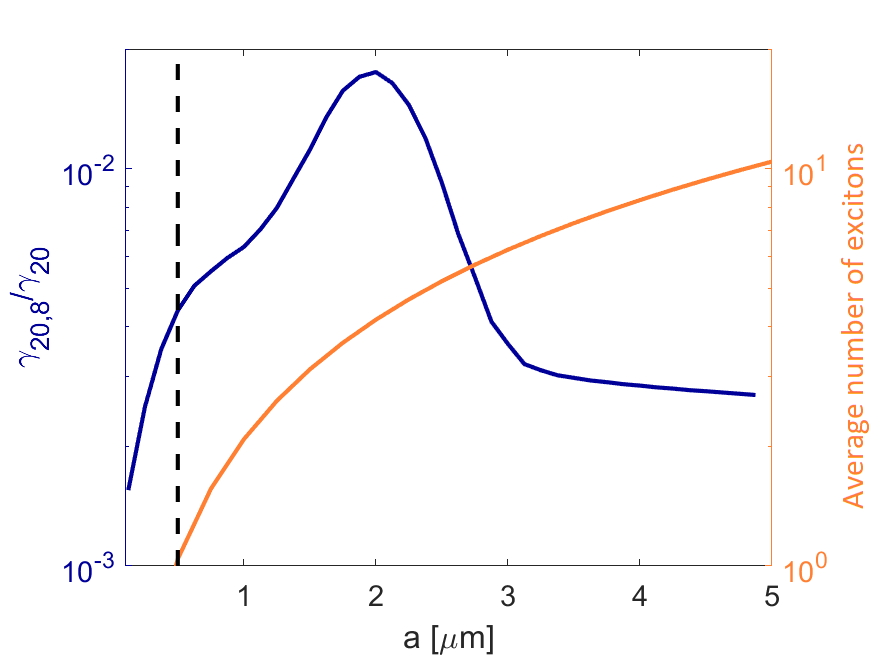}
\caption{The ratio of the transition rates $\gamma_{20,8}/\gamma_{20}$ and the number of excitons inside a single groove as a function of $a$.}\label{Fig_7}
\end{figure}

One interesting possibility is to take advantage of the large blockade volume of excitons characterized by large principal quantum number $n$ to facilitate single photon emission. Khazali \textit{et al} proposed a system where only a single exciton could fit inside a Cu$_2$O nanocrystal \cite{Khazali17}, which resulted in an emission of a single photon when it decayed. Here, we propose a microwave analogue of this system by using extremely narrow grooves. Specifically, let's consider a system characterized by $h=\lambda/4$ and $a<5$ $\mu$m. The thickness of the system in $y$ axis is $b=10$ $\mu$m. The excited state is $n=20$ and the system is matched to the transition $n=20 \rightarrow n=8$, which is characterized by a wavelength $\lambda \approx 0.98$ mm. The results are shown on  Fig. \ref{Fig_7}.
Due to the large dipole moment of the $n=20 \rightarrow n=8$ transition and longer lifetime of $n=20$ exciton, the ratio $\gamma_{20,8}/\gamma_{20}$ is slightly larger than in the last system, reaching peak efficiency of $1.7\%$ for $a = 2$ $\mu$m. As mentioned in \cite{Rusina10}, the capability of the system to focus the EM field is limited by the skin depth of the metal, which in the THz range is of the order of 120 nm. One can see that in this particular example, the transition rate sharply decreases when $a<500$ nm (dashed line), which is four times the skin depth. At the point of peak efficiency, the average number of excitons that fit inside the groove is of the order of 10. It should be mentioned that since the exciton size is comparable to the groove width $a$, a more thorough analysis involving confined excitonic states \cite{Hamid,Konzelmann} and their interaction with metallic surfaces \cite{Kolhoff16} would be needed to fully describe the system. Nevertheless, it is clear that with sufficient reduction of the groove width, one can easily fabricate a system where a single exciton serves as a source of single, microwave photons. 

\section{Conclusions}
We have proposed a copper-based plasmonic system that facilitates microwave to optical frequency conversion with the use of excitonic states in Cu$_2$O. The described scheme takes advantage of the capability of surface plasmons to amplify and focus the field in subwavelength areas, which facilitates the enhancement of transition probabilities with the Purcell effect. We focus on the microwave frequency transitions between highly excited levels of Rydberg excitons in Cu$_2$O. It is shown that with proper geometry, one can use the so-called spoof plasmons to reduce the plasmon frequency to microwave range, allowing for efficient plasmon-Rydberg exciton coupling. With optimal geometry, the amplification of the probability of specific microwave transitions is of the order of $10^3$, with values up to $10^5$ being feasible. This, in turn, provides relatively large conversion efficiency of over $1\%$ \cite{Lambert20}. Finally, the possibility of single-photon emission with the use of Rydberg blockade is investigated. The system geometry (corrugated surface) and materials (copper and copper oxide) are characterized by low cost and ease of fabrication and due to the large number of available excitonic levels, one can choose transitions within a large range of wavelengths (0.1 - 10 mm) \cite{maser}.

\section{Acknowledgments}
Support from the National Science Centre, Poland (NCN), project
Miniatura, 2022/06/X/ST3/01162, is acknowledged.

\appendix
\section{FDTD method}
For numerical calculations of the field distribution around the plasmonic structure, we use the Finite-Difference Time-Domain (FDTD) method \cite{Yee}, which is based directly on Maxwell's equations and provides a high flexibility and accuracy in modelling the propagation of electromagnetic waves. To reduce computational demands, we assume that the system is two-dimensional, e.g. the system depicted in Fig. \ref{Fig_1} is much larger than wavelength in $y$ axis. The computation domain is divided by a rectangular grid with a single cell size $\Delta x=10$ $\mu$m and the electromagnetic field components corresponding to waves propagating on $xz$ plane ($\vec{E}=[E_x,0,E_z]$ and $\vec{H}=[0,H_y,0]$. are calculated with evolution formulas derived from Maxwell's equations. Specifically,
\begin{eqnarray}
\frac{\partial H_y}{\partial t}&= \frac{1}{\mu_0}\left(\frac{\partial E_x}{\partial z}-\frac{\partial E_z}{\partial x} \right),\\
\frac{\partial E_x}{\partial t}&= \frac{1}{\epsilon_0}\left(\frac{\partial H_y}{\partial z}-\frac{\partial P_x}{\partial t} +j_x\right),\\
\frac{\partial E_z}{\partial t}&= \frac{1}{\epsilon_0}\left(\frac{\partial E_x}{\partial z}-\frac{\partial P_z}{\partial t}+j_z \right),
\end{eqnarray}
where $j_x$,$j_z$ are components of current density, $\epsilon_0$, $\mu_0$ are the vacuum permittivity and permeability, respectively and $P$ is medium polarization. The above equations are discretized with a fixed time step $\Delta t$ and rearranged to obtain time derivatives of the $E_x$, $E_z$, $H_y$ fields, which are used to calculate next field values from current ones iteratively. In the calculations, a unit normalization is used so that $\epsilon_0=\mu_0=c=1$, $\Delta_x=1$, $\Delta t=0.5$. The time and spatial step satisfy the Courant stability criterion \cite{Taflove}
\begin{equation}
\frac{c\Delta t}{\Delta x}<\frac{1}{\sqrt{2}}.
\end{equation}
For the considered system, the wavelength $\lambda=1$ mm is equal to 100 spatial steps and thus the numerical frequency is
\begin{equation}
\omega = \frac{2\pi c}{\lambda} \approx 0.063.
\end{equation}

The dispersive properties of media are included with the use of auxiliary differential equation (ADE) approach \cite{Okada}, where the time evolution of polarization vector $\vec{P} = [P_x, 0, P_z]$ is described by a second-order partial differential equation
\begin{equation}\label{e_ADE}
\ddot{\vec{P}}+\Gamma_j \dot{\vec{P}}+\omega_j=\frac{\epsilon_0 f_j}{{\epsilon_\infty}}\vec{E},
\end{equation}
with a constant permittivity $\epsilon_\infty$ and a set of fitted oscillator terms $j=1,2,3...$ characterized by oscillator strength $f_j$, damping $\Gamma_j$ and resonant frequency $\omega_j$. For the particular example of the system considered here, the surface plasmon is generated by a monochromatic wave and thus it has a relatively narrow spectrum. In such a case, the dispersion of copper oxide is negligible and thus it can be characterized by a constant $\epsilon_\infty=7.5$. For copper, we use Drude model described by Eq. (\ref{e_ADE}), with a single oscillatory term characterized by $f_1=24.6875$, $\Gamma_1=0.000916$, $\omega_1=0$. In the frequency domain, the copper permittivity is given by
\begin{equation}
\epsilon(\omega)=1-\frac{f_1}{\omega^2+ i\Gamma_1\omega}
\end{equation}
For the above mentioned frequency $\omega=0.063$, one obtains $\epsilon(\omega)=-6218 + 90i$, which is consistent with ref. \cite{Tukmakova20}.

The calculation is performed for a system depicted in Fig. \ref{Fig_2}, with numerical grid size of 400x200 points (4 mm and 2 mm respectively). A line source of radiation (single point on xz plane) is used to excite surface plasmons. After steady plasmon field amplitude is reached, the source is turned off. Then, the energy density is calculated according to Eq. (\ref{eq:density}). The decay of total field energy in time is used to estimate the $Q$ factor of the system.

\end{document}